\documentclass[conference]{ieeeconf}
\IEEEoverridecommandlockouts
\usepackage{cite}
\usepackage[utf8]{inputenc}
\usepackage{graphicx}
\usepackage{amsmath}
\usepackage[version=4]{mhchem}
\usepackage{siunitx}
\usepackage{longtable,tabularx}
\usepackage{subcaption}
\usepackage{epstopdf}
\usepackage{tabularx}
\usepackage{booktabs} % \**rule table line package
\usepackage{multirow}
\usepackage{arydshln} % dashline package
\usepackage{url}
\usepackage{float}
\usepackage{acronym}
\usepackage{hyperref} % makes references hyperlinks
\usepackage{textcomp} % Symbol Expansion Package
\usepackage{tabularx,multirow,array}
\usepackage{threeparttable}
\usepackage{subcaption}
\usepackage{xcolor} % For text coloring
\usepackage{amssymb} % For \checkmark symbol

\def\BibTeX{{\rm B\kern-.05em{\sc i\kern-.025em b}\kern-.08em
    T\kern-.1667em\lower.7ex\hbox{E}\kern-.125emX}}

\renewcommand{\footnoterule}{%
  \kern -3pt
  \hrule width \columnwidth height 1pt
  \kern 2pt
}

\begin{document}

\title{Learning-Aided Control of Robotic Tether-Net with Maneuverable Nodes to Capture Large Space Debris}
% {\footnotesize \textsuperscript{*}Note: Sub-titles are not captured in Xplore and
% should not be used}
% \thanks{Identify applicable funding agency here. If none, delete this.}
% }

\makeatletter
\newcommand{\linebreakand}{%
  \end{@IEEEauthorhalign}
  \hfill\mbox{}\par
  \mbox{}\hfill\begin{@IEEEauthorhalign}
}
\makeatother

% \author{Achira Boonrath$^*$ \thanks{$^*$ Ph.D. Student, Department of Mechanical and Aerospace Engineering, AIAA Student Member}, Feng Liu $^*$, Eleonora M. Botta$^\dagger$ \thanks{$^\dagger$ Assistant Professor, Mechanical and Aerospace Engineering, AIAA Senior Member}, Souma Chowdhury$^\ddag$  \thanks{$^\ddag$ Associate Professor, Mechanical and Aerospace Engineering, AIAA Senior Member, Corr. author}}

% \and
% \IEEEauthorblockN{6\textsuperscript{th} Given Name Surname}
% \IEEEauthorblockA{\textit{dept. name of organization (of Aff.)} \\
% \textit{name of the organization (of Aff.)}\\
% City, Country \\
% email address or ORCID}
\author{Achira Boonrath$^{1\dag}$, Feng Liu$^{1\dag}$, Eleonora M. Botta$^{1}$, and Souma Chowdhury$^{1}$% <-this % stops a space
\thanks{$^1$Department of Mechanical and Aerospace Engineering, University at Buffalo, Buffalo, NY 14260, USA}
\thanks{\dag These authors contributed equally}
\thanks{{\tt\small \{achirabo, fliu23, ebotta, soumacho\} @buffalo.edu}}%
\thanks{*This work is supported under the CMMI Award numbered 2128578 from the National Science Foundation (NSF). The author's opinions, findings, and conclusions or recommendations expressed in this material do not necessarily reflect the views of the National Science Foundation. The authors would also like to thank CM Labs Simulations for providing licenses for the Vortex Studio simulation framework. }
}

\maketitle
% \begingroup\renewcommand\thefootnote{\textsection}
% \footnotetext{Equal contribution}
% \endgroup

\begin{abstract}
Maneuverable tether-net systems launched from an unmanned spacecraft offer a promising solution for the active removal of large space debris. Guaranteeing the successful capture of such space debris is dependent on the ability to reliably maneuver the tether-net system -- a flexible, many-DoF (thus complex) system -- for a wide range of launch scenarios. Here, scenarios are defined by the relative location of the debris with respect to the chaser spacecraft. This paper represents and solves this problem as a hierarchically decentralized implementation of robotic trajectory planning and control and demonstrates the effectiveness of the approach when applied to two different tether-net systems, with 4 and 8 maneuverable units (MUs), respectively. Reinforcement learning (policy gradient) is used to design the centralized trajectory planner that, based on the relative location of the target debris at the launch of the net, computes the final aiming positions of each MU, from which their trajectory can be derived. Each MU then seeks to follow its assigned trajectory by using a decentralized PID controller that outputs the MU's thrust vector and is informed by noisy sensor feedback (for realism) of its relative location. System performance is assessed in terms of capture success and overall fuel consumption by the MUs. Reward shaping and surrogate models are used to respectively guide and speed up the RL process. Simulation-based experiments show that this approach allows the successful capture of debris at fuel costs that are notably lower than nominal baselines, including in scenarios where the debris is significantly off-centered compared to the approaching chaser spacecraft.
\end{abstract}

\newacro{ADR}{Active Debris Removal}
\newacro{RL}{Reinforcement Learning}

% \begin{IEEEkeywords}
% PID control, reinforcement learning, robotic tether net, space debris removal 
% \end{IEEEkeywords}


\section{Introduction}
 Active Debris Removal (ADR) has emerged as a promising solution to counteract the increasing threats posed by space debris to satellites and other assets orbiting the Earth. Among various proposed ADR methodologies, using a tether-net system possesses a high potential for success due to the system's relatively low weight and ability to capture massive rotating debris from a secure distance \cite{LEDKOV2022100858, BottaPhdThesis,shanADRreview, chen2022analysis}. In recent years, many researchers have performed simulation-based and experiment-based analyses on net systems' deployment and capture dynamics \cite{benvenuto2015dynamics, medina2017validation, golebiowski2016validated, botta2019simulation, endo2020study, hou2021dynamic}. Most past research on the topic focuses on passive tether-net systems, which function through the ejection of masses attached to the perimeter or edges of nets toward target debris. The trajectory of such systems cannot be altered once the net has been launched from the chaser. To further improve the capabilities of tether-net systems, employing maneuverable space nets enhances the mission's flexibility and dependability in capturing uncooperative space debris. Addressing this need, Huang et al. \cite{huang2015dynamics,ZhangFan2018Scoa, ZhaoYakun2022DCPD} introduced the concept of the Tethered Space Net Robot (TSNR) system. The TSNR configuration comprises a net connected via threads to Maneuverable Units (MUs). Compared to a purely passive net, using MUs allows the system to increase flexibility through its ability to alter its trajectory after launching from the chaser. 

Within the literature, the setpoints for the TSNR system to fly towards are predetermined before the initialization of the net deployment \cite{huang2015dynamics,ZhangFan2018Scoa}. This method assumes that the chaser spacecraft is able to always perfectly approach the debris based on a pre-determined proximity approach trajectory. Not only is this unlikely in practice, but also differing debris rotational rates (that may not be known a priori) could demand different approach states. Hence, an approach that allows determining setpoints or optimal net deployment on the fly, based on the estimation of the debris' state relative to the chaser, is premised to provide a significantly more generalizable solution to debris capture. We posit that Reinforcement Learning (RL) can be used to train such generalized policies across various approach scenarios. Therefore, our work introduces a RL-guided autonomous maneuverable net system, in which MUs are guided to provide successful capture of space debris at minimal fuel costs, based on the state of the debris relative to the chaser. It is assumed that the MUs are equipped with cold-gas thrusters commanded by PID controllers. 

In addition to the 4-MU TSNR system design from prior work by the authors \cite{jsr}, \cite{liu2023learning}, this work introduces a TSNR system with 8 MUs. The alternate design explores the feasibility of using maneuvers of docking among the MUs to replace the closing mechanism (that is otherwise necessary \cite{botta2019simulation}) in ensuring capture success, and the potential benefits thereof. The system components are shown in Fig. \ref{fig:sket} for the 8-MU net, which is identical to the 4-MU net, except for the 4 additional MUs on the net sides and the lack of a separate closing mechanism.

\begin{figure}[h]
    \centering
    \includegraphics[width=0.42\textwidth]{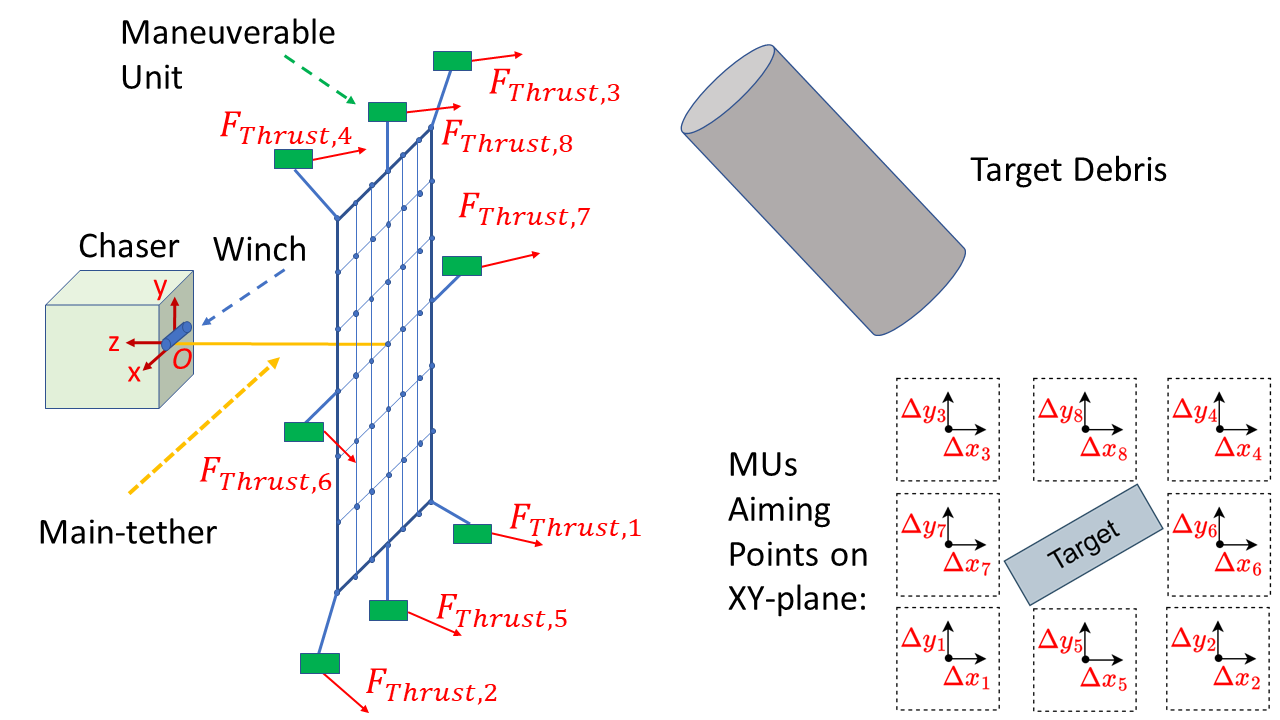}
    \caption{Representation of the 8-MU Tether-net System}
    \label{fig:sket}
    \vspace{-15pt}
\end{figure}

In our previous work, RL \cite{arl} and neuroevolution \cite{neruoev,behjat-agent-IEEE-TAI-2023} techniques were respectively used to perform open-loop thrust control of a maneuverable net system \cite{liu2023learning} and purely mechanical net ejection and closing control \cite{zeng2022}, both with the goal to maximize the success rate of debris capture. Instead, in this paper, we assume that each MU is able to fly a trajectory using a PID controller to reach a desired end location, which we call the \textit{aiming point}, as shown in Fig. \ref{fig:sket}. This aiming point is determined by a policy model trained using RL, given the debris' relative location with respect to the chaser. To update the policy, the Capture Quality Index (CQI), the number of locked pairs of closing nodes of the net, the total fuel consumption of the MUs, and the net mouth area are used as the simulated evaluation metrics to compute a reward function. To circumvent the cost of simulating the net/debris contact mechanics, and therefore speed up the RL process, a surrogate model is trained to predict the capture quality based on the position and velocities of selected points on the net right before the net mouth starts closing. 
% \begin{figure}[h]
%     \centering
%     \includegraphics[width=0.5\textwidth]{images/flowchart.png}
%     \caption{Proposed Policy Learning Process} 
%     \label{fig:Network}
% \end{figure}

% \begin{figure}[h]
%     \centering
%     % \includegraphics[width=0.3\textwidth]{images/Aimpoint.png}
%     \includegraphics[width=0.2\textwidth]{images/aimpoint_high_res.jpg}
%     \caption{Aiming Point for the MUs} 
%     \label{fig:aimpoint}
% \end{figure}

The primary contributions of this paper and its outline can be summarized as: \textbf{1)} We present a new concept of operation for tether-net systems based on autonomously maneuverable MUs in tandem with or without a closing mechanism, depending on the number of MUs (Sec. \ref{DynamicsofDMSNModel}). \textbf{2)} We present a new closed-loop (PID) controller for trajectory tracking by each MU under sensing uncertainties (Sec. \ref{DynamicsofDMSNModel}). \textbf{3)} We present a simulation environment that combines the net's deployment dynamics model with a surrogate model of the capture success, leading to an 8-fold speed-up of the RL process; \textbf{4)} We formulate the trajectory planning problem for each MU as a time-encoded linear path that is defined by its initial position and aiming point (on the plane parallel to the inertial X-Y plane and intersecting the center of mass of the debris or target), with the aiming point computed by solving a Markov Decision Process (MDP) using the policy gradient method (Sec. \ref{LearningtheOptimalApproachingPolicies}). Performance of the resulting tether-net maneuver process is compared to nominal cases in Sec. \ref{results}. 

\section{Tether-Net: ConOps and Simulation}\label{DynamicsofDMSNModel}

\subsection{Concept of Operation (ConOps)}

In the first stage of the net-based ADR operation, the chaser spacecraft is launched and rendezvous with the target debris in orbit. Once in close proximity, the relative displacement between the target debris and the chaser is determined through a LiDAR system installed on the chaser; this concept was demonstrated as a part of the RemoveDEBRIS in-orbit demonstration mission \cite{aglietti2020removedebris, chabot2017vision}. Before the launch of the TNSR, considering the debris state, a semi-decentralized RL-based trajectory planner computes and relays information regarding the optimal desired position at the end of deployment for each MU (aiming points). To launch the net, the MUs utilize their propulsion capabilities; this eliminates the necessity for an ejection mechanism, thus removing either gas-based or spring-based launchers required for passive net designs \cite{BottaPhdThesis, botta2019simulation}. Employing closed-loop control, each MU applies thrusts to follow a designated trajectory until a start-of-closing condition is reached (see Sec. \ref{sec:SimDescr}). To secure the debris, the net mouth is then closed by a closing mechanism in the case of the 4-MU system, or purely by the thrusters in the case of the 8-MU system. No human intervention is involved from the moment of debris position determination to the end of the capture. After the debris is secured, the chaser satellite is assumed to bring it to a disposal orbit; the post-capture phase is not simulated in this work, but is the topic of concurrent work \cite{field2023relative}.
\vspace{-5pt}
\subsection{Simulation Description}
\label{sec:SimDescr}

The modeling of the deployment of the TSNR and the closing of the net around the debris are performed using a formerly assembled simulator in the multibody dynamics simulation framework Vortex Studio. The chaser satellite is modeled as a cubic rigid body with a side length of 1.5 m and a mass of 1600 kg. During the capture process, the chaser spacecraft is assumed to be floating without control to enable the examination of motion induced by tension in the main tether. The main tether links the net’s central knot to a winch on the chaser, which is designed to spool freely during deployment and to remain locked during the capture phase \cite{BottaPhdThesis, botta2019simulation}. The target debris used here is the $\sim$9000 kg second stage of the Zenit-2 rocket, which is considered to be one of the most dangerous debris currently orbiting the Earth \cite{MCKNIGHT2021282, wiedemann2012cost}.

The lumped-parameter modeling method represents a flexible net within the simulation framework \cite{BottaPhdThesis, botta2019simulation}. The process lumps the mass of the net into many small spherical rigid bodies (called \textit{nodes}); the nodes of the net are linked together into a mesh via spring-damper elements that are unable to withstand compression \cite{BottaPhdThesis, botta2019simulation}. The MUs are modeled as small rectangular prism-shaped rigid bodies, each possessing a dimension of 0.1 m $\times$ 0.1 m $\times$ 0.2 m and a total weight of 2.5 kg. The simulator leverages a scaled-box friction model to compute the frictional forces during contact (approximating Coulomb’s friction model); the normal contact forces are calculated by continuous compliant contact forces modeling. The parameters of the nets studied within this paper are the same as in  \cite{liu2023learning}. For more comprehensive insights into the modeling of the net and of contact dynamics, readers are encouraged to refer to prior work by Botta et al. \cite{BottaPhdThesis,botta2019simulation, botta2016simulation}.

Two system variants are considered for the TNSR. The first is characterized by a 4-MU layout, also studied in previous work \cite{liu2023learning}. In this case, a closing mechanism is utilized to close the mouth of the net around debris after contact; it consists of threads that are interlaced around the net's perimeter and of winches placed in each MU. The winches reel the threads into the MUs at the net closing activation time, thus pulling the nodes together. The maximum number of locked pairs, i.e., the number of pairs of adjacent nodes on the closing mechanism that are locked together, is  $\max(N_\text{L})=12$. 
%This mechanism had been used on the net tested within the RemoveDEBRIS in-orbit demonstration mission \cite{aglietti2020removedebris, axthelm2017net}. 

As an alternative design, a net that utilizes the maneuvering capabilities of the MUs to close the net mouth is introduced. Since in previous work by Botta et al. \cite{botta2019simulation} it was observed that solely relying on the four corner masses to close the net's mouth might fail to maintain a secure hold on the targeted debris, four additional MUs are introduced to assist in closing the net perimeter.
% since they allow for more control over the net perimeter shape during the capture phase. 
Each of the four additional MUs is attached to the center of a side of the net via a thread. At the start-of-closing time, the MUs fly towards a set closing position and dock with each other. To simulate the docking process, when the adjacent MUs are within a distance of 0.5 m, \textit{distance joint} constraints with a 0 m nominal length are engaged within the simulation: this ensures that the net mouth remains closed and capture is maintained. In this 8-MU design, the maximum value of locked pairs $\max(N_\text{L})=8$ should be achieved for an ideal capture. 
In the 8-MU design, each of the MUs could be simpler and smaller due to the absence of a winch, compensating for the increase in total system mass. Additionally, the new design benefits from increased redundancy since the failure of 1 MU is foreseen to have a smaller overall effect on net deployment. 

\subsection{Thrust Control}
To control the motion of the MUs, thrust forces, produced by cold-gas thrusters and defined as $\mathbf{F}_{\text{T},i}$, are employed. For state estimation, each MU is assumed to be equipped with an inertial measurement unit  with a sampling rate of 20 Hz (based on sensors currently employed within CubeSats \cite{colagrossi2023effective, sanders2013pushing, tullino2017testing}). To represent sensing or measurement errors, random noises within 3$\sigma$ bounds of $\pm$ 0.1 m and $\pm$ 0.1 m/s are sampled from a Gaussian distribution and respectively added to the positions and velocities obtained from the simulation. These slightly noisy state values are utilized by each controller, with the actuation assumed to be perfect. 

During deployment, PID controllers compute the thrust needed to bring each MU to its desired position. At each command timestep, the desired position of $i$-th MU is: 
\begin{equation}
\vspace{-3pt}
	\mathbf{r}_{d,i}(t)=\mathbf{r}_{0,i} + \frac{t}{t_{\text{final}}}(\mathbf{r}_{\text{final},i}-\mathbf{r}_{0,i})\\
 \label{eq:fuelConsumed}
 \vspace{-3pt}
\end{equation}
\noindent where $\mathbf{r}_{0,i}$, $t_{\text{final}}$, and $\mathbf{r}_{\text{final},i}$ are the initial position of the $i$-th MU, the desired duration of deployment, and the desired position of the MU at the end of the deployment, respectively. The position $\mathbf{r}_{\text{final},i}$ is the position of the aiming point (located on the plane parallel to the inertial X-Y plane and passing through the debris' center of mass), which is the point the $i$-th MU must aim for. Note that in practice, the closing mechanism is usually triggered slightly before the MUs reach their aiming points since the trigger event is defined as the time when the net's center of mass is 2.5 m away from the debris' center of mass \cite{liu2023learning}. The RL-trained policy determines these aiming positions $\mathbf{r}_{\text{final},i}$ for each MU based on the state of the debris with respect to the point of net-launch at the instant when the tether net is launched. In this work, $t_{\text{final}}$ is a user-prescribed net's deployment time value, set as 25 s.

The PID controller assumes that thrust can be independently controlled in the $x$, $y$, and $z$ directions. As such, the control force for the $i$-th MU is computed as:
% \begin{align} 
% 	\label{eq:Fx}
% 	{F}_{T,i,$x$}&=K_P(x_i-x_{d,i}) - K_D \dot{$x$}_i + K_I\int_{0}^{t} (x_{d,i} -x_i) \,dt\\
% 	\label{eq:Fy}
% 	{F}_{T,i,$y$}&=K_P(y_i-y_{d,i}) - K_D \dot{$y$}_i + K_I\int_{0}^{t} (y_{d,i} -y_i) \,dt\\
% 	\label{eq:Fz}
% 	{F}_{T,i,$z$}&=K_P(z_i-z_{d,i}) - K_D \dot{$z$}_i + K_I\int_{0}^{t} (z_{d,i} -z_i) \,dt
% \end{align}
\begin{equation}
\mathbf{F}_{\text{T},i}=K_P(\mathbf{r}_{d,i}-\mathbf{r}_i) - K_D \mathbf{\dot{r}}_i + K_I\int_{0}^{t} (\mathbf{r}_{d,i} -\mathbf{r}_i) \, d\tau
% \vspace{-15pt}
\end{equation}
\noindent where $K_P$, $K_D$, and $K_I$ are the proportional, derivative, and integral gains, respectively. Variables $\mathbf{r}_i$ and $\mathbf{\dot{r}}_i$ are the position and inertial velocity of the $i$-th MU at any time. Activation of the thrusters takes place at $t = 0$ s. For the 4-MU net, the thrusters deactivate once the predetermined distance between the center of mass of the net and the debris is achieved. The magnitude of the thrust in each direction is limited to 5.1 N, chosen so that the total maximum thrust available for each MU is (approx. 8.9 N) the same as in previous work \cite{liu2023learning}. Each thruster has a command frequency of 20 Hz \cite{silik2020control}. In between each command signal, a zero-order hold is utilized. The structure of the PID controller is shown in the dashed box in Fig. \ref{fig:RL_flow}.
%where $K_P$,  is the proportional gain, $K_D$ is the derivative gain, $K_I$ is the integral gain
%, and $\mathbf{r}_i$ and $\mathbf{\dot{r}}_i$ are the position and inertial velocity of the $i$-th MU at any time. Activation of the thrusters takes place at $t = 0$ s. For the 4-MUs net, the thrusters deactivate once the predetermined distance between the center of mass of the net and the debris is achieved. The magnitude of the thrust in each direction is limited to 5.1 N, chosen so that the total maximum thrust available for each MU is (approx. 8.9 N) the same as in previous work \cite{liu2023learning}. Each thruster has a command frequency of 20 Hz \cite{silik2020control}. In between each command signal, a zero-order hold is utilized. The structure of the PID controller is shown in the dashed box in Fig. \ref{fig:RL_flow}.
%
% \begin{figure}[h!]
% \vspace{-0.2cm}
%     \centering
%     \includegraphics[width=0.4\textwidth]{images/PID2.png}
%     \caption{PID Controller Design} 
%     \label{fig:pid}
% \end{figure}
%

To compute the fuel consumption of the thruster on each MU, a specific impulse $I_{sp} =60 $ s is chosen, based on existing cold-gas thruster technology \cite{yost2023state}. For the $i$-th MU, the total fuel utilized throughout the maneuver is:
\begin{equation}
	m_{\text{f},i}=\int_{0}^{t_{\text{end}}} \frac{\bar{F}_{\text{T},i}}{(g_0 I_{sp})} \,dt \\
 \label{eq:fuelConsumed}
\end{equation}
\noindent where $g_0$ is the gravitational acceleration at Earth's sea level (i.e., 9.81 m/s$^2$), and $\bar{F}_{\text{T},i}$ is the magnitude of the saturated thrust force. Variable $t_{\text{end}}$ indicates the final simulation time, set to be 15.0 s after the start of the closing of the net's mouth (achieved using a winch in the case of the 4-MU system or the MUs themselves in the case of the 8-MU system). Therefore, Eq. \eqref{eq:fuelConsumed} accounts for both deployment and closure. 
% \vspace{-5pt}

\subsection{Capture Quality Index and Locked Pairs}

A quantitative measure of the quality of debris capture, called CQI, is utilized for this work. This provides a measure of the ability of the TSNR to wrap around and hold onto the debris without examining simulation graphics \cite{CQI_Original, barnes}. The CQI compares the geometries of the net's Convex Hull (CH) and of the debris, in addition to accounting for the distance between their centers of mass. The CQI is thus defined as:
\begin{equation}
    I_{\text{CQI},n} = 0.1\frac{|V_n-V_D|}{V_{\text{D}}}+0.1\frac{|S_n-S_{\text{D}}|}{S_{\text{D}}} +0.8\frac{|q_n|}{L_c}
    \label{eq:cqiSafe}
\end{equation}
\noindent At the $n$-th time-step, $V_n$, $V_{\text{D}}$, $S_n$, and $S_{\text{D}}$ respectively represent the net's CH volume, the debris's volume, the net's CH surface area, and the debris's surface area; $q_n$ and $L_c$ respectively represent the distance between the net's and debris's centers of mass and the debris's characteristic length. The CQI can distinguish between successful and unsuccessful captures, as shown in  \cite{barnes}. In this work, however, the CQI is complemented by the number of locked pairs, which provides information on the fact that capture will not be lost. In the current framework, a successful capture is defined to have a settled CQI (i.e., CQI value 15 s after closing activation) lower than 2.5 and a number of locked pairs greater than a certain threshold (8 for the 4-MU net and 6 for the 8-MU net). For the second stage of the Zenit-2 rocket, the geometric propertiesare $V_{\text{D}}=159.9$ m$^3$, $S_{\text{D}}=59.9$ m$^2$, and $L_c=1.95$ m.
\vspace{-4pt}

\subsection{Surrogate Modeling to Predict Capture Status}
With the high-fidelity simulator, computing time could become excessive for the simulation to be directly used in optimization or learning processes. Over a set of 50 random successful capture scenarios (running in parallel on a 16-Core Processor Windows workstation with 64 GB RAM, where on average, deployment took 25.2 s of physical time and capture was simulated for 15 s of physical time), the average computing time taken to simulate the deployment phase and the capture phase were observed to be respectively 97.1 s and 206.7 s. Hence, the computing time is dominated by the simulation of the capture process, which involves net closing and net/debris contact. This motivated the use of surrogate models to substitute the capture simulation. Through numerical experiments, recurrent neural network (RNN) modeling was observed to outperform other multi-variate regression or surrogate models (e.g., multi-layer perceptron, random forest regression). Hence, RNN is used in this work as a surrogate model to predict the CQI and the number of locked pairs in the capture phase.

% Hence, it was decided that substituting the capture phase of the simulation (in Vortex) with a surrogate model could provide 5-10 times speed up in evaluating net deployment and debris capture episodes.

The inputs to the RNN model are the positions and velocities of MUs and of 165 nodes on the net, relative to the debris's center of mass. The 165 net nodes (out of 533 total) form three loops on the net, which are used as a reference to represent the net's configuration in space. High-fidelity simulation stops at the start-of-closing time. Subsequently, the RNN reads the position and velocity (input) data of the MUs and of selected nodes and predicts the final CQI and number of locked pairs at time $t_{\text{end}}$. The latter are used in the RL framework to compute the reward function as will be explained in Eq. \eqref{eq:reward}. 

The RNN network for the 4-MU case study consists of 990 neurons in the input layer, 500 and 300 neurons in two consecutive hidden layers, and two neurons in the output layer, which provide the values of the CQI and number of locked pairs. The RNN network for the 8-MU case study has a similar structure, except that the input layer comprises 1014 neurons to account for the states of more MUs. The RNN model for the 4-MU system was trained with 2540 scenarios, and that for the 8-MU system was trained with 2790 scenarios, both using a 0.00001 learning rate and 500 epochs. The models were then tested over 200 unseen scenarios, where the accuracy of computing whether a capture is successful (refer to constraints in Eq. \eqref{eq:reward} given by the RNN predictions) was found to be over 98\%.
%
%The performance of the RNN models over the training data is listed in Table \ref{tab:rnn_result}, which show that the Mean Squared Error (MSE) of RNN predictions for the number of locked pairs is low. While the successful capture scenarios and the RNN model predictions are quite accurate, the overall MSE skews higher due to greater errors incurred in CQI predictions for failed cases. 
To compensate for the RNN's prediction error (which was observed to be high in failed capture cases) and promote robust learning for scenarios with CQI$<$20, the prediction error is represented by a Gaussian distribution. This serves as the noise model from which the CQI is sampled during the RL process to compute rewards.

\vspace{-0.15cm}

\section{Formulation of the Learning task\vspace{-0.12cm}}\label{LearningtheOptimalApproachingPolicies}

The objective of RL in this paper is to find the aiming points in different launch scenarios, where scenarios (state space) are defined in terms of the $x$, $y$, and $z$ coordinates of the debris' center of mass.
% w.r.t. the point of net-launch on the chaser. 
These aiming points define the optimal positions to be achieved by the MUs when the closing process is initiated. Optimal aiming points should minimize the fuel cost (due to the MUs thrusting to maneuver the net) while ensuring that debris is successfully captured, with the latter determined in terms of settled CQI and number of locked pairs.

Proximal Policy Optimization (PPO) \cite{ppo}, provided by stable-baselines3 \cite{stable-baselines}, is used to perform the RL process. The learning framework is shown in Fig. \ref{fig:RL_flow}, comprising the following major elements: 1) the policy model being trained by PPO, 2) the high-fidelity simulation, including the PID controller, to model the net deployment, and 3) the RNN based surrogate model to predict the measures of capture success.
\begin{figure}[h!]
    \centering
    \includegraphics[width=0.39\textwidth]{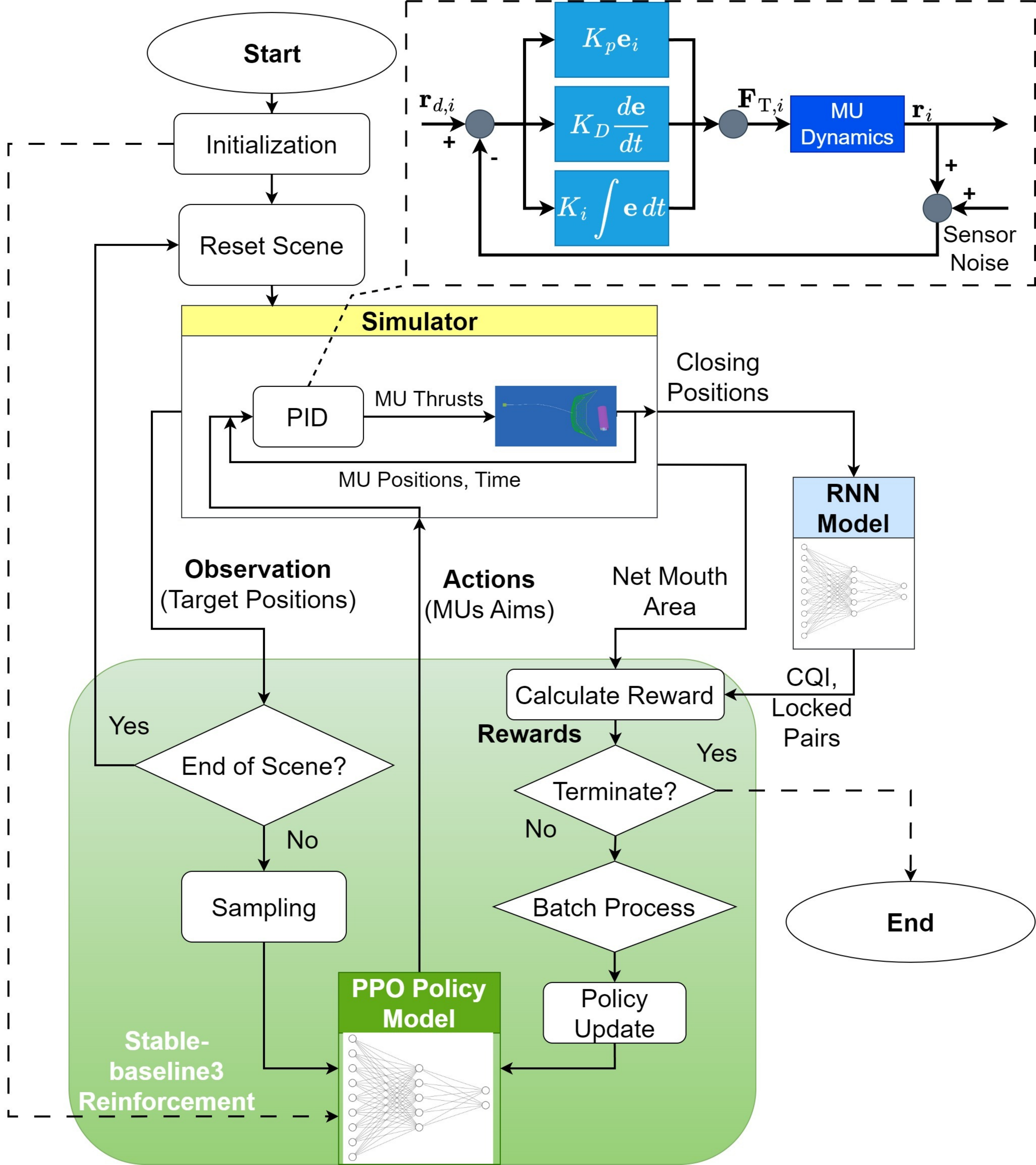}
    \caption{Computational Framework for Reinforcement Learning of Aiming and Control of the MUs} 
    \label{fig:RL_flow}
%    \vspace{-0.8cm}
\end{figure}
% RL models the system as a Markov Decision Process (MDP) \cite{mdp}, and the bounds states and actions of the MDP are listed in Table \ref{tab:state_action}.

The Markov Decision Process over which this RL process operates can be described in terms of the state space, the action space, the state transition model, and the reward function.

\textbf{State Space}: The state space comprises the position vector of the center of mass of the debris
% w.r.t. the point of net-launch on the chaser, measured 
at the instant of net launch, and is expressed as $S=\left[x_{\text{D}},y_{\text{D}},z_{\text{D}}\right]$. Note that it is a one step experience here, that is, a single action is taken given this state in every episode simulating a net launch scenario. The state transition model is stochastic since noise is added to the simulation to mimic sensor imperfections and to the RNN model to account for uncertainty in the predicted CQI and locked pairs. 
%its uncertainty. 

% A set of nominal position for each MU ($x_{\text{nom}, i}$, $y_{\text{nom}, i}$, and $z_{\text{nom}, i}$) based on debris position ($x_{t}$, $y_{t}$, and $z_{\text{D}}$) can be seen in in Table \ref{tab:nomMU}.
\textbf{Action Space}: The actions of the RL policy are the offsets of the aiming points for each MU relative to nominal aiming points (defined as a baseline) $\Delta x_i$ and $\Delta y_i$, as shown in Fig. \ref{fig:sket}. 
% The nominal positions are defined on the $x$-$y$ plane passing through the debris' center of mass, as shown in Fig. \ref{fig:sket}. 
The position of the nominal point for the $i$-th MU can be expressed as $\left(x_{\text{nom}, i},y_{\text{nom}, i},z_{\text{D}}\right)$; the nominal positions are defined in Table \ref{tab:nomMU}.
% where $x_{\text{nom}, i},y_{\text{nom}, i}$ represents an offset from the center of mass of debris ($x_{\text{D}},y_{\text{D}},z_{\text{D}}$) . 
The nominal aiming point allows the net to be flattened out at the approximate time the closing of the net mouth starts. Both the 4-MU and 8-MU designs use the same nominal coordinates for MU 1 to MU 4. The RL-trained policy model is tasked to compute where the optimal aiming points must be placed, compared to the nominal points. The actual aiming point for any $i$-th MU is thus given by $\mathbf{r}_{\text{final},i}=(x_{\text{nom},i}+\Delta x_{i})\mathbf{\hat{i}}+(y_{\text{nom},i}+\Delta y_{i})\mathbf{\hat{j}}+z_{\text{D},i}\mathbf{\hat{k}}$. Hence, the actions computed by the RL policy for each MU are $A_i = \left[\Delta x_{i},\Delta y_{i}\right]$. Both nominal and policy-based $z$-coordinates for all the aiming points are set to be the same to prevent the net configuration from becoming too asymmetrical.

\textbf{Reward}: The reward function is shown in Eq. \eqref{eq:reward}, which considers multiple factors. Firstly, it rewards the policy if the MUs spend less fuel in total and are able to expand the net mouth wider before the closing mechanism is triggered. The reward function penalizes the policy if the MUs fail to capture the target debris, which is determined based on the value of settled CQI and the number of locked pairs. In Eq. \eqref{eq:reward}, \(\Phi\) is the RL policy. Here, \(b_{\text{fuel}} = 1 - \frac{m_\text{f}}{m_\text{fmax}}\) to promote higher reward when the fuel consumption is lower, where \(m_\text{f}\) is the total fuel consumption of all MUs and \(m_\text{fmax}\) is the reference maximum total fuel consumption, which is set based on prior observations; \(w\) is a positive weighting parameter, set at 1 in the 4-MU case and 1.5 in the 8-MU case; \(A_{\text{c}}\) is the net mouth area when the closing mechanism activates, and \(A_{\text{max}}\) is the maximum net mouth area; \(I^*_{\text{CQI}}\) and \(N_{\text{L}}\) are respectively the CQI and number of locked pairs 15 s after the net mouth closing is triggered; \(N_{\text{D}}\) is the minimum number of locked pairs for successful capture (8 for the 4-MU case and 6 for the 8-MU case).
\begin{table}[h]
\vspace{-5pt}
    \centering
    \begin{threeparttable}
    \caption{Bounds of RL States and Actions}
    \label{tab:state_action}
    \begin{tabular}{lllll}
        \toprule
        & Variables & Bounds & Step Size & Unit\\
        \midrule
        \multirow{3}{*}{State} & $x_{\text{D}}$ & -9 to 9 & 0.1 & m\\
                               & $y_{\text{D}}$ & -9 to 9 & 0.1 & m\\
                               & $z_{\text{D}}$ & -60 to -40 & 0.1 & m\\
        \midrule
        \multirow{2}{*}{Action} & $\Delta x_{i, i=1,2, ..., N}$  & -5 to 5 & 0.1 & m\\
                                & $\Delta y_{i, i=1,2, ..., N}$  & -5 to 5 & 0.1 & m\\
        \bottomrule 
    \end{tabular}
    \begin{tablenotes}
        \item[*] $N$ is the number of MUs.
    \end{tablenotes}
    \end{threeparttable}
    \vspace{-15pt}
\end{table}
\begin{table*}[h]
	\begin{center}
	\caption{\label{tab:nomMU} Nominal End-of-Deployment Coordinates for Each MU}
		\begin{tabular}{|l|l|l|l|l|l|l|l|l|l|}
        \hline
		Coordinates, m & MU 1 & MU 2 & MU 3 & MU 4 & MU 5 & MU 6 & MU 7 & MU 8\\
        \hline
        $x_{\text{nom}}$ &$x_{\text{D}}$-12.00 & $x_{\text{D}}$+12.00 & $x_{\text{D}}$-12.00 & $x_{\text{D}}$-12.00 & $x_{\text{D}}$ &  $x_{\text{D}}$+11.70 & $x_{\text{D}}$-11.71 &    $x_{\text{D}}$\\
        \hline
        $y_{\text{nom}}$ &$y_{\text{D}}$-12.00 & $y_{\text{D}}$-12.00 & $y_{\text{D}}$+12.00 & $y_{\text{D}}$+12.00 & $y_{\text{D}}$-11.71  & $y_{\text{D}}$ & $y_{\text{D}}$& $y_{\text{D}}$-11.71 \\
        \hline
        % $z_{\text{nom}}$ & $z_{\text{D}}$ & $z_{\text{D}}$ &   $z_{\text{D}}$  &  $z_{\text{D}}$ & $z_{\text{D}}$  & $z_{\text{D}}$  &  $z_{\text{D}}$  & $z_{\text{D}}$ \\
        %  \hline
		\end{tabular}
	\end{center}
 \vspace{-10pt}
\end{table*}
\begin{align}
\label{eq:reward}
    \max_{\Phi}\quad & R = b_{\text{mouth}} + p_{\text{CQI}} + p_{\text{NL}} + b_{\text{\text{end}}} \\
    \text{where: }\quad 
    &b_{\text{mouth}} = {A_c}/{A_\text{max}}\notag\\
    &p_{\text{CQI}} =
        \begin{cases}
         -\ln((I^*_{\text{CQI}}-2.5)^2+1), & \text{if } I^*_{\text{CQI}} > 2.5 \\
         0,              & \text{otherwise}
        \end{cases}\notag\\
    & p_{\text{NL}} =
        \begin{cases}
         -\ln((N_\text{L}-N_{\text{D}})^2+1), & \text{if } N_\text{L} < N_{\text{L,max}} \\
         0,              & \text{otherwise}
        \end{cases}\notag\\
    & b_{\text{\text{end}}} =
        \begin{cases}
         w \, b_{\text{fuel}}, & \text{if } I^*_{\text{CQI}}\leq2.5 \wedge N_\text{L}\geq N_{\text{L,max}}\\
         0,              & \text{otherwise}
        \end{cases}\notag            
\end{align}
% Where \(\Phi\) is the RL policy, \(b_{\text{fuel}} = 1 - \frac{m_\text{f}}{m_\text{fmax}}\) indicates that the reward is higher when the fuel consumption is lower, \(w\) is a positive parameter used to tune the weight of the fuel reward, \(m_\text{f}\) is the total fuel consumption of all MUs and \(m_\text{fmax}\) is the maximum total fuel consumption of all MUs observed in previous simulations. \(A_c\) is the net mouth area when the closing mechanism activates. \(A_{max}\) is the maximum net mouth area. \(I^*_{CQI}\) is CQI 15.0 s after the start of the net mouth closing. \(N_\text{L}\) is the number of locked pairs 15.0 s after the start of the net mouth closing. \(N_t\) is the minimum number of locked pairs for successful capture (8 for 4 MUs case and 6 for 8 MUs).

\section{Simulation Results and Analysis}\label{results}
\subsection{Net Deployment PID Control: Results}
A Routh table was constructed to determine the stable range of gain values to select controller gains. Values of $K_P=10.0$, $K_I=6.0$, and $K_D=6.0$ were picked from the determined range and used for each MU. The performance of the controller at tracking a given deployment trajectory is demonstrated for the 8-MU case in Fig. \ref{fig:4muError}, which displays the magnitude of the position error over time, computed as:
\begin{equation}
    e(t)_{i} = ||\mathbf{r}_{d,i}(t) - \mathbf{r}_{i}(t) || 
    \label{eq:cqiSafe}
\end{equation}
\noindent It can be seen that the error for all MUs is minimal after 10 seconds post-launch, demonstrating the ability of the controller to track the given position input most of the time throughout the deployment. A slight increase in the position error is observed starting at approximately 24 s, which corresponds to when the net becomes almost flattened out. To improve the controller performance in future work, tuning the PID gains may reduce the time required for the MUs to reach a near-zero position error and the tracking ability towards the end of the deployment. In Fig. \ref{fig:4muError}, the net-to-debris's centers of mass distance is also shown to steadily decrease during the deployment, confirming the approach of net towards the debris. 
% further showing the utility of the PID controller in ensuring stable deployment of the net.

\begin{figure}[h]
\vspace{-10pt}
     \centering
         \includegraphics[width=0.40\textwidth]{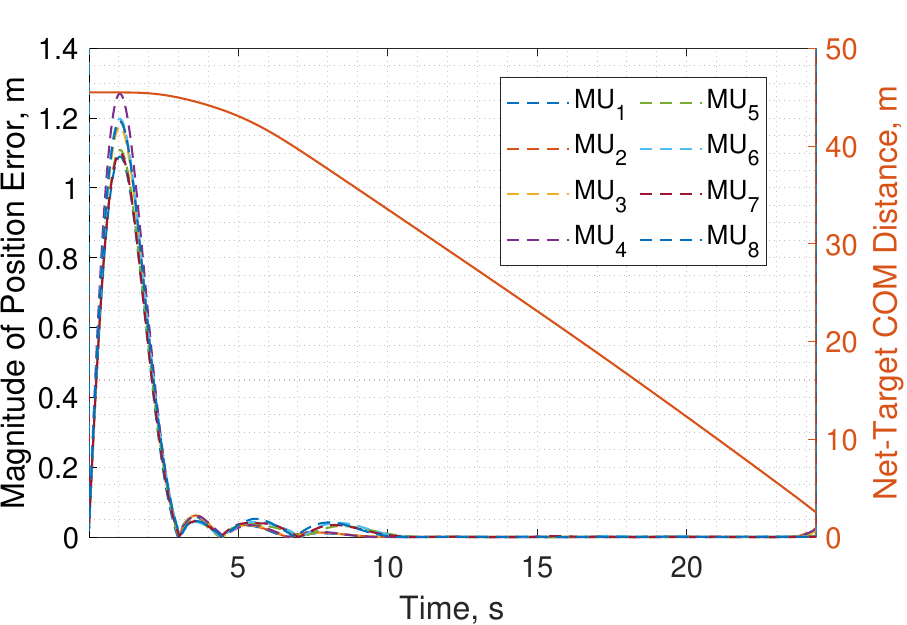}
         \caption{L2-Norm of MUs position error (left y-axis) and distance of net from the debris (right y-axis) over time}
         \label{fig:4muError}
\vspace{-15pt}
\end{figure}

\subsection{Aiming Point Planning via RL: Results}

A Windows workstation with an AMD Ryzen 9 5950X 16-core Processor and 64 GB RAM was utilized for the RL training for both the 4-MU and 8-MU designs. The training was performed with 32 episodes in parallel, with a mini-batch size of 64 and a learning rate of 0.001. For the 4-MU design, 11360 episodes were used in the learning process, which took 14 hours. For the 8-MU case, 11360 episodes were used, which took 10.5 hours. 

The corresponding reward averaged over 32 consecutive episodes is reported to show the training history in Fig. \ref{fig:reward_mu}. For the chosen reward function, the reward for each episode takes values between $-2.0$ and $2.5$. For each episode, a negative reward corresponds to a failed capture caused by completely missing the debris or by failure to activate the closing procedure. A reward between 0 and 1 corresponds to a capture where the CQI threshold or number of locked pairs threshold is not reached. A reward greater than 1 corresponds to a successful capture. From the training history in Fig. \ref{fig:reward_mu}, it can be seen that in the 4-MU and 8-MU cases, the reward converges to a value of approximately 1.5 and 2.0, respectively. This observation confirms the effectiveness of the designed RL framework and settings. 
\begin{figure}[t]
    \vspace{-15pt}
    \centering
    \includegraphics[width=0.42\textwidth]{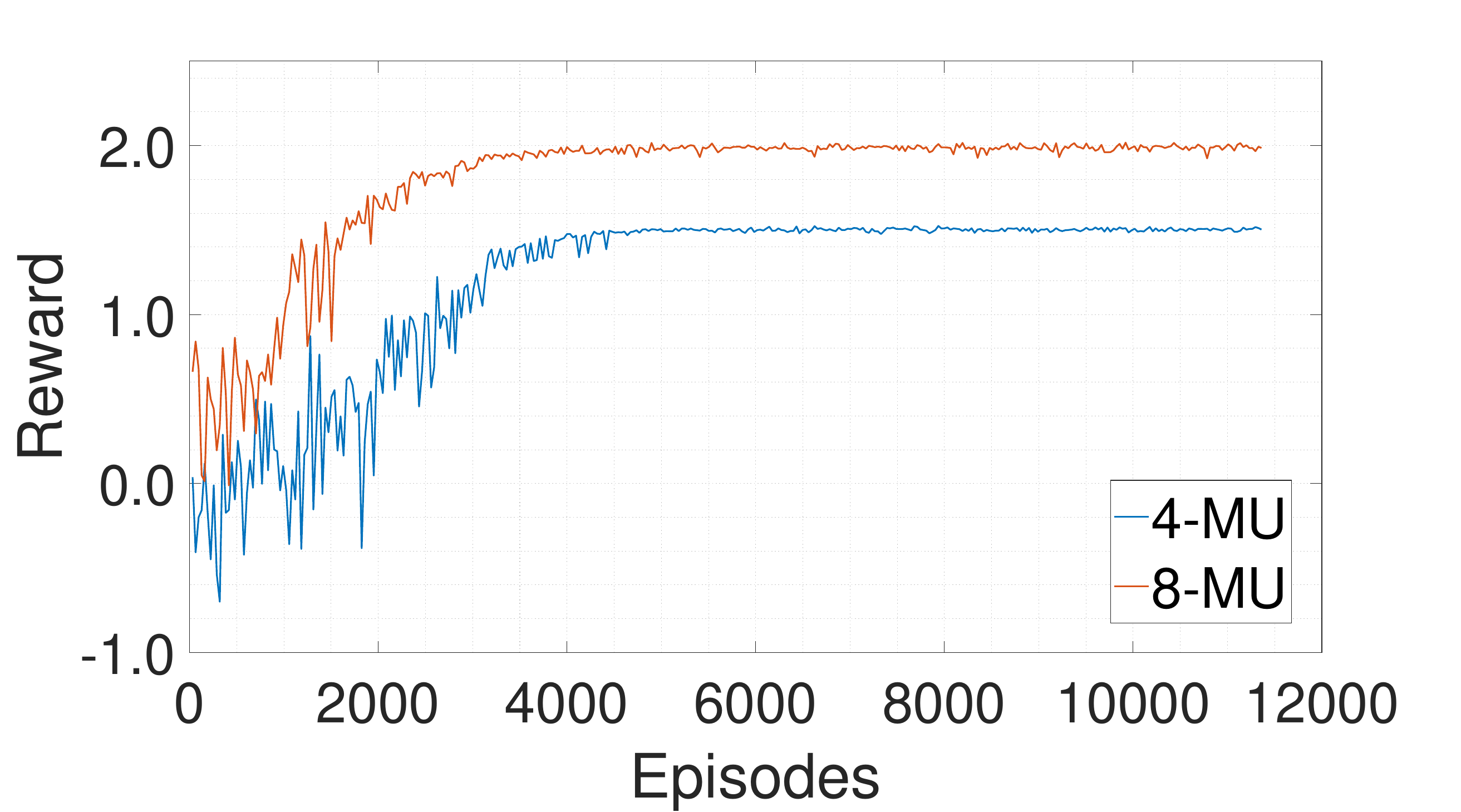}
    \caption{Training History: Reward Averaged Over 32 Episodes}
    \label{fig:reward_mu}
   \vspace{-15pt}
\end{figure}

Figure \ref{fig:opt50} shows snapshots of key moments in an example scenario, with a notable offset of the debris of $x_{\text{D}}=8.9$ m, $y_{\text{D}}=6.2$ m, and $z_{\text{D}}=-44.0$ m, for (a) the 4-MU and (b) 8-MU systems. The left images display the systems in an early stage of the deployment, while the right images display the systems approximately 10 s after the closing of net mouth was initiated, demonstrating successful capture.

%For this simulation, the total fuel consumption of all of the MUs combined is ... kg.
\begin{figure}[h]
\centering
% Group for first two subplots
\begin{minipage}{.5\linewidth}
    \centering
    \begin{subfigure}{.4\linewidth}
      \includegraphics[width=0.95\linewidth]{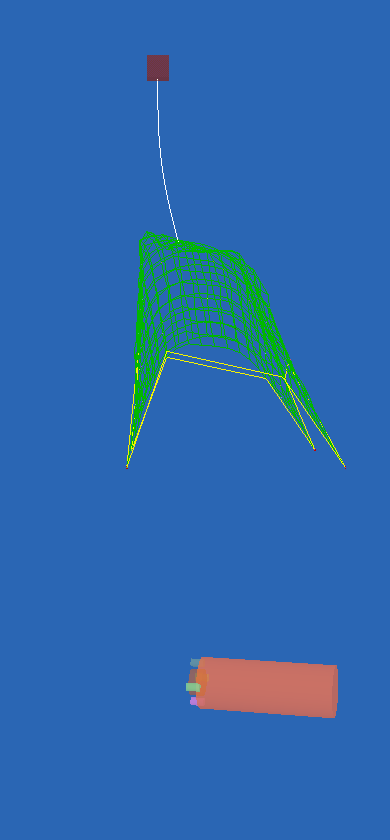}
     \label{fig:sub1}
    \end{subfigure}%
    \begin{subfigure}{.4\linewidth}
      \includegraphics[width=0.95\linewidth]{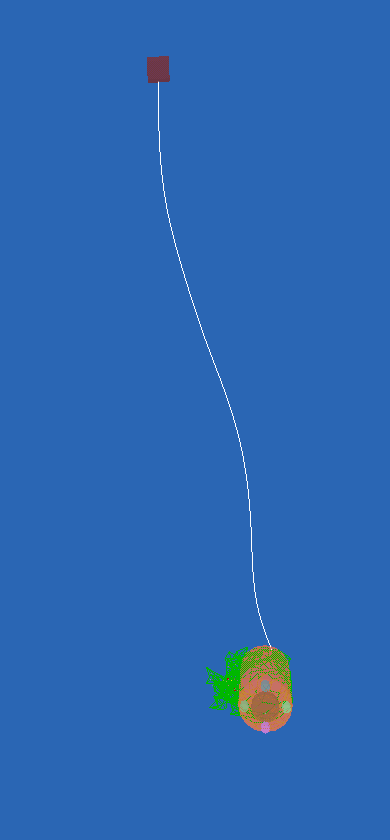}
      \label{fig:sub2}
    \end{subfigure}
    \subcaption{4-MU at 15 s and 35 s}
\end{minipage}%
% Group for second two subplots
\begin{minipage}{.5\linewidth}
    \centering
    \begin{subfigure}{.4\linewidth}
      \includegraphics[width=0.95\linewidth]{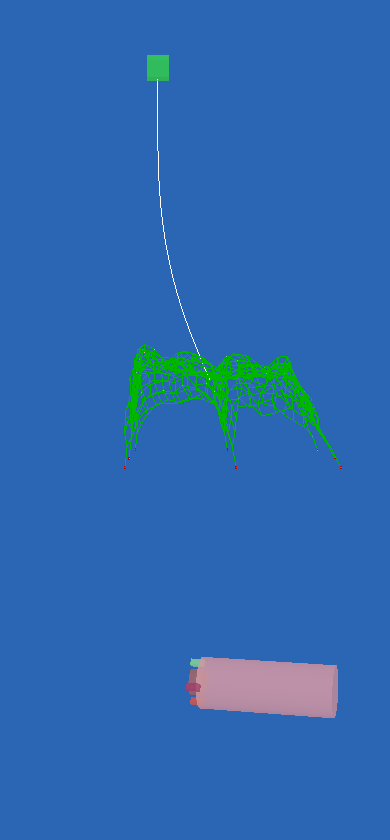}
      \label{fig:sub3}
    \end{subfigure}%
    \begin{subfigure}{.4\linewidth}
      \includegraphics[width=0.95\linewidth]{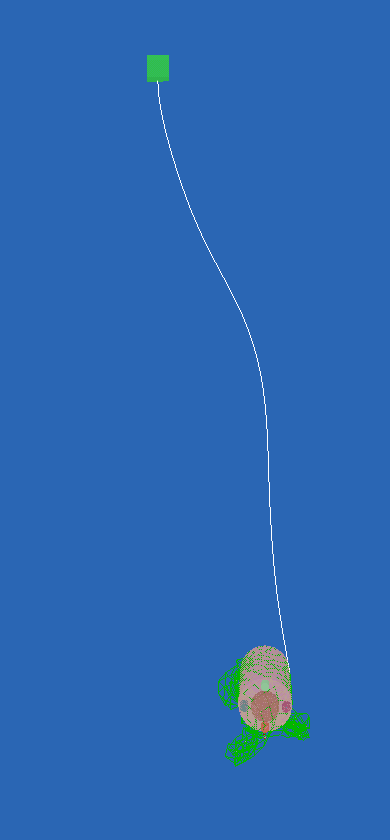}
      \label{fig:sub4}
    \end{subfigure}
    \subcaption{8-MU at 15 s and 35 s}
\end{minipage}
\caption{Simulation Screenshots Utilizing RL Guidance}
\label{fig:opt50}
\vspace{-10pt}
\end{figure}

To evaluate the performance of the RL policy model, 100 random unseen test scenarios (with different locations of the debris with respect to the chaser) are used. Evaluation is performed using the high fidelity simulation from net launch to the end of capture. The capture success rate over the testing samples is observed to be 100\%. This is an important improvement over the Case 1 RL policy in our earlier work \cite{liu2023learning}, which considered a direct open-loop control of the thrusts of each MU in the 4-MU case and reported 88\% success over a more limited range of debris/chaser offsets. 

Figure \ref{fig:fuel_4mu} shows the comparison of the fuel consumption by the RL-trained systems vs. using the nominal aiming points (see Table \ref{tab:nomMU}). In particular, the reduction in fuel mass from the nominal setting to the RL cases (i.e., $m_{\text{f},\text{nom}} - m_{\text{f},\text{RL}}$, the greater, the better) is plotted as boxplots for the same scenarios. On average, the RL-based policies reduce fuel consumption by 11\% in the 4-MU case, and by 1.8\% in the 8-MU case, in comparison to their respective nominal aiming baselines. In addition, we observed that, across the test scenarios, the average fuel consumption of each MU by the 4-MU system was higher (at 0.028 kg) than that by the 8-MU system (at 0.021 kg). The rate of capture success remained 100\% across the 100 test scenarios for both the RL-trained 4-MU and 8-MU cases. Collectively, these observations provide evidence of the feasibility of performing effective closing directly with MUs without the use of a winch mechanism. 

\begin{figure}[t]
\vspace{-15pt}
         \centering
         \includegraphics[width=0.4\textwidth]{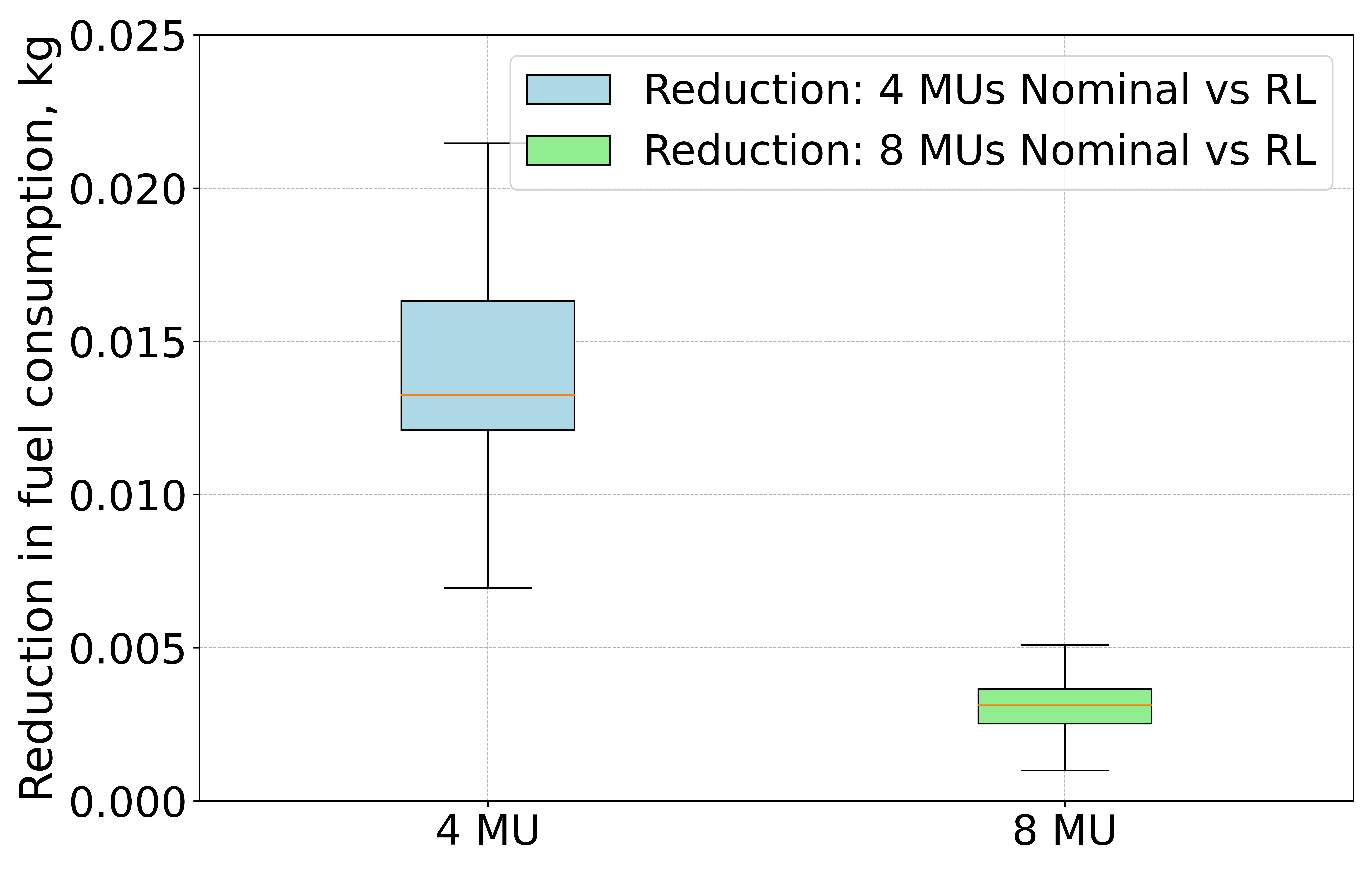}
         \caption{Comparison of Total Fuel Consumption Reduction for 4- and 8-MU Cases}
         \label{fig:fuel_4mu}
\vspace{-15pt}
\end{figure}
% Similar to the 4 MUs case study, the 8 MUs RL policy was also tested against 100 nominal cases utilizing a set of random target positions. The difference between the simulations utilizing the RL policy and the nominal case can be seen in Fig. \ref{fig:fuel_4mu} with the right boxplot. On average, the MU flight path, as determined by the policy, has a 1.8\% lower fuel consumption compared to the nominal path. 

% The capture success rate in the 100 tests is again 100\%. This demonstrates the feasibility of using only the MUs to close and secure the net mouth on a rotating target at various positions relative to the chaser. Additionally, while the average fuel consumption for each MU for over 100 simulations with the 4 MU design is 0.028 kg, the average fuel consumption for the 100 simulations utilizing the 8 MU design for each MU is lower at 0.021 kg. Thus, with a view to a single MU, the 8 MU design is more efficient than the 4 MU design. This advantage allows for the MUs in the 8 MUs design to be smaller due to both the absence of a winch and lower required fuel, including the amount needed to close the net mouth, for operation.

\section{Conclusion}
This paper proposed learning-aided PID control of the maneuverable units (MUs) of a tether-net launched from a chaser spacecraft for the capture of large space debris. More specifically, RL is used to learn the aiming points for the MUs as a function of the location of the debris relative to the chaser. Based on that, a trajectory is derived and flown thanks to PID controllers, which decide the thrusts on the MUs. To accelerate the RL process, an RNN-based surrogate model is used to replace the more time-consuming portion of the simulation, and predict capture quality metrics. Tested on two different net designs (i.e., a 4-MU system with winch-based closing, and a 8-MU system that uses the MUs to close), the proposed framework is found to provide significant reduction in total fuel costs compared to the baseline, while ensuring 100\% capture success over unseen test scenarios. 
% Interestingly, the total fuel costs incurred by the 8-MU system were observed to be lower on average compared to the 4-MU system. 
At present, the control framework utilized in this work ignores the attitude dynamics of the MUs, which is an important concern in actual spaceflight. Consideration of position and attitude control in the future will allow us to also reliably expand the action space to include both aiming position and velocity. This, along with increased realism in considering vision-based sensing of the debris state and accounting for its rotation rates, will enable us to generate further insights regarding the effectiveness of autonomous tether-net systems for space debris capture.

\bibliographystyle{IEEEtran}
\bibliography{sample} % Path to your .bib file

\end{document}